# Fabrication of free-standing Pt nanowires for use as thermal anemometry probes in turbulence measurements

Hai Le-The[1,2,4], Christian Küchler[3,4], Albert van den Berg[2,4], Eberhard Bodenschatz[3,4], Detlef Lohse[1,4], Dominik Krug[1,4]

[1]Physics of Fluids Group, MESA+ Institute, University of Twente, Enschede 7522 NB, The Netherlands

[2]BIOS Lab-on-a-Chip Group, MESA+ Institute, University of Twente, Enschede 7522 NB, The Netherlands

[3]Max Planck Institute for Dynamics and Self-Organization, 37077 Göttingen, Germany

[4]Max Planck-University of Twente Center for Complex Fluid Dynamics

Correspondence:   Hai Le-The (h.lethe@utwente.nl)

Dominik Krug (d.j.krug@utwente.nl)

**ABSTRACT**

We report a robust fabrication method for patterning free-standing Pt nanowires for the use as thermal anemometry probes for small-scale turbulence measurements. Using e-beam lithography, high aspect ratio Pt nanowires (~300 nm width, ~70 µm length, ~100 nm thickness) were patterned on the surface of oxidized silicon (Si) wafers. Combining wet etching processes with dry etching processes, these Pt nanowires have been successfully released free-standing between two silicon dioxide ($SiO_2$) beams supported on Si cantilevers. Moreover, the unique design of the bridge holding the device allowed to release the device gently without damaging the Pt nanowires. The total fabrication time was minimized by restricting the use of e-beam lithography to the patterning of the Pt nanowires while standard photolithography was employed for other parts of the devices. We demonstrate that the fabricated sensors are suitable for turbulence measurements when operated in a constant-current mode. A robust calibration between output voltage and fluid velocity was established over the velocity range from 0.5 m s$^{-1}$ to 5 m s$^{-1}$ in an $SF_6$ atmosphere at a pressure of 2 bar and a temperature of 21°C. The sensing signal from the nanowires showed negligible drift over a period of several hours. Moreover, we confirmed that the nanowires are able to withstand high dynamic pressures by testing them in air at room temperature velocities up to 55 m s$^{-1}$.

**INTRODUCTION**

Even today, fully resolved measurements of flow velocities in highly turbulent flows remain highly challenging. The difficulty is best illustrated by considering the non-dimensional Reynolds number (*Re*), which measures turbulence intensity by relating the magnitudes of inertial and viscous forces acting in the flow. Accessing high *Re* flows experimentally is important from a practical perspective as many engineering applications, such as the boundary layers on the hulls of ships and planes or flow problems in wind farms, fall into this regime. Moreover, measurements in high *Re* flows are also highly relevant to foster and to validate our theoretical understanding of turbulence.

A hallmark of turbulence is the fact that 'eddying motions', i.e. seemingly random velocity fluctuations, across a wide range of scales contribute to the evolution of the flow. The range of spatial scales with *Re* as $L/\eta \sim Re^{3/4}$, which renders the measurement challenge obvious[1]. If the largest scale *L* is fixed, *e.g.* by the size of the lab facilities, high *Re* can only be reached if the smallest scale $\eta$ (the so-called Kolmogorov scale) is decreased in size. Typical sizes of $\eta$ – and consequently the spatial resolution requirements – are in the order of micrometers. Also the temporal resolution is essential to resolve the short turnover timescales of such small eddies[2,3]. Especially in cases where flow structures are advected past the probe by a strong mean flow, such as in investigations of turbulent boundary layers, frequency requirements can reach orders of 100 kHz[4].

To date, the best resolution and bandwidth characteristics for measuring turbulent velocity fluctuations are achieved using 'hotwire anemometry' (HWA), which is a proven technique with a long history[5–8]. Its measurement principle is based on the velocity dependent convective cooling of a heated wire element (with wire diameter *d*) that is placed in the fluid. The time varying cooling leads to changes in the wire electrical resistance and thus to a voltage signal in the attached electrical circuit, which can be calibrated to yield a fluid velocity measurement. The effective sensor size in HWA is given by the length ($\ell$) of the wire. However, $\ell$ cannot be decreased arbitrarily. This is because a shorter wire length also increases the portion of the heat that leaves the wire via end-conduction, which is unwanted and detrimental to the measurement. This issue can only be overcome if shorter wires are also made thinner. Traditionally a minimum aspect ratio $\ell/d \leq 200$ has been used[9], while more recently Hultmark et al.[10] provided a refinement of this criterion. The conventional wire filaments with the best performance characteristics are produced from so-called "Wollaston wires" (thin Pt wires clad in silver) by etching away part of the silver jacket. The sensing element is then formed by the exposed platinum (Pt) wire for which minimum diameters of about 1 µm can be achieved in this way. Pushing beyond this limit has proven very difficult despite significant efforts. For example, Willmarth & Sharma produced wires with the length of 50 µm using a Wollaston wire of 0.5 µm in diameter[11]. However, given the relatively low aspect ratio, the performance of this design was hampered by end-conduction effects. Ligrani & Bradshaw[9] stuck to an aspect ratio of approximately 200 when designing the wires with a diameter of 0.625 µm, but with a minimum of 125 µm the resulting wire length was still rather large. The need to decrease sensor sizes below this limitation initiated a push towards nanofabrication techniques. Early efforts by Löfdahl et al.[12] yielded only moderate improvements as their probes featured a large sensing area. Jiang et al.[13] employed microelectromechanical systems (MEMS) techniques to fabricate a poly-silicon thermal anemometry probe but the very good spatial resolution came at the price of significant end-conduction losses in their case. End-conduction is also a problem for the multi-component hot-wire probes (50 µm × 6 µm × 2.7 µm) fabricated by Chen et al.[14]. Moreover, being fixed to a wall, these sensors are also not suitable for conventional turbulence measurements.

More recently, the development of the nanoscale thermal anemometry probe, termed NSTAP[15–19], provided a breakthrough towards unprecedented small-scale resolution. Some noteworthy later developments, such as a micro fabricated multi-array probe that provides access to the full velocity gradient tensor[20], or a specialized hotwire sensor for measurements in cryogenic helium[21] have been reported since. For completeness, it should also be mentioned that MEMS techniques have been employed to fabricate small-scale cantilevers for flow measurements[22,23], but the measurement principle (beam deflection) is different in those cases. In terms of sensor size the NSTAP remains the state of the art to date. The production process of the NSTAP combines standard photolithography with a series of dry-etching and wet-etching processes. The sensing element consists of a Pt wire, which is approximately 100 nm thick, while its width is still 1 μm. The latter is a limitation of the photolithography process but in part also a choice in order to enhance the convective heat transfer from the wire[16]. Note also that for a variant of the NSTAP, the q-NSTAP reported by Fan *et al.*[18], electron-beam lithography is employed. This reduces the width of the wire to between 600 nm and 800 nm. However, with a length of only 10 μm, the q-NSTAP is designed to measure humidity and is not suited for anemometry. Even with these reduced wire dimensions, the authors report issues regarding the structural integrity of the sensor due to internal stresses originating from wet etching of silicon dioxide ($SiO_2$) to release the wire.

Despite these efforts, measurement resolution remains the bottleneck for investigations of very high *Re* turbulence in a well-controlled lab environment. In an effort to push the envelope on this, we report a robust method for the fabrication of free-standing Pt nanowires here. These novel wires feature a reduced cross section (300 nm width, 100 nm thickness) compared to existing sensors. A lower cross section offers several advantages. On the one hand, it will allow to reduce the effective sensing length while keeping the aspect ratio constraint and thereby limiting conduction losses. Note that with a length of 70 μm we made a rather conservative choice in the design reported here, since as far as the fabrication and robustness are concerned, longer wires are more challenging. On the other hand, reducing the cross section also reduces the thermal inertia of the sensor, which will lead to a better frequency response. Moreover, by approaching an aspect ratio of 1 between width and thickness of the wire, we expect to eliminate spurious angular sensitivity of the measured velocity signal. In this paper, we describe how by combining e-beam lithography with wet etching processes and dry etching processes, Pt nanowires have been successfully fabricated free-standing between two silicon dioxide (SiO$_2$) beams supported on Si cantilevers. We further confirm that the fabricated nanowires are capable and sufficiently robust to measure the velocity of turbulent flows even at large fluid densities. We tested this in the Variable Density Turbulence Tunnel (VDTT) with pressurized SF$_6$ as working fluid as well as in an air flow with velocities up to 55 m s$^{-1}$ without damaging the wires.

**RESULTS AND DISCUSSION**

Figure 1 presents an overview over the processing sequence for the fabrication process of a device featuring a free-standing Pt nanowire. Further details on the dimensions of the structure are provided in the Supplementary information (Figure S1). We elaborate on individual fabrication steps in the following. Further details and the specific processing parameters employed are provided in the Materials and Methods section.

Free-standing Pt nanowires

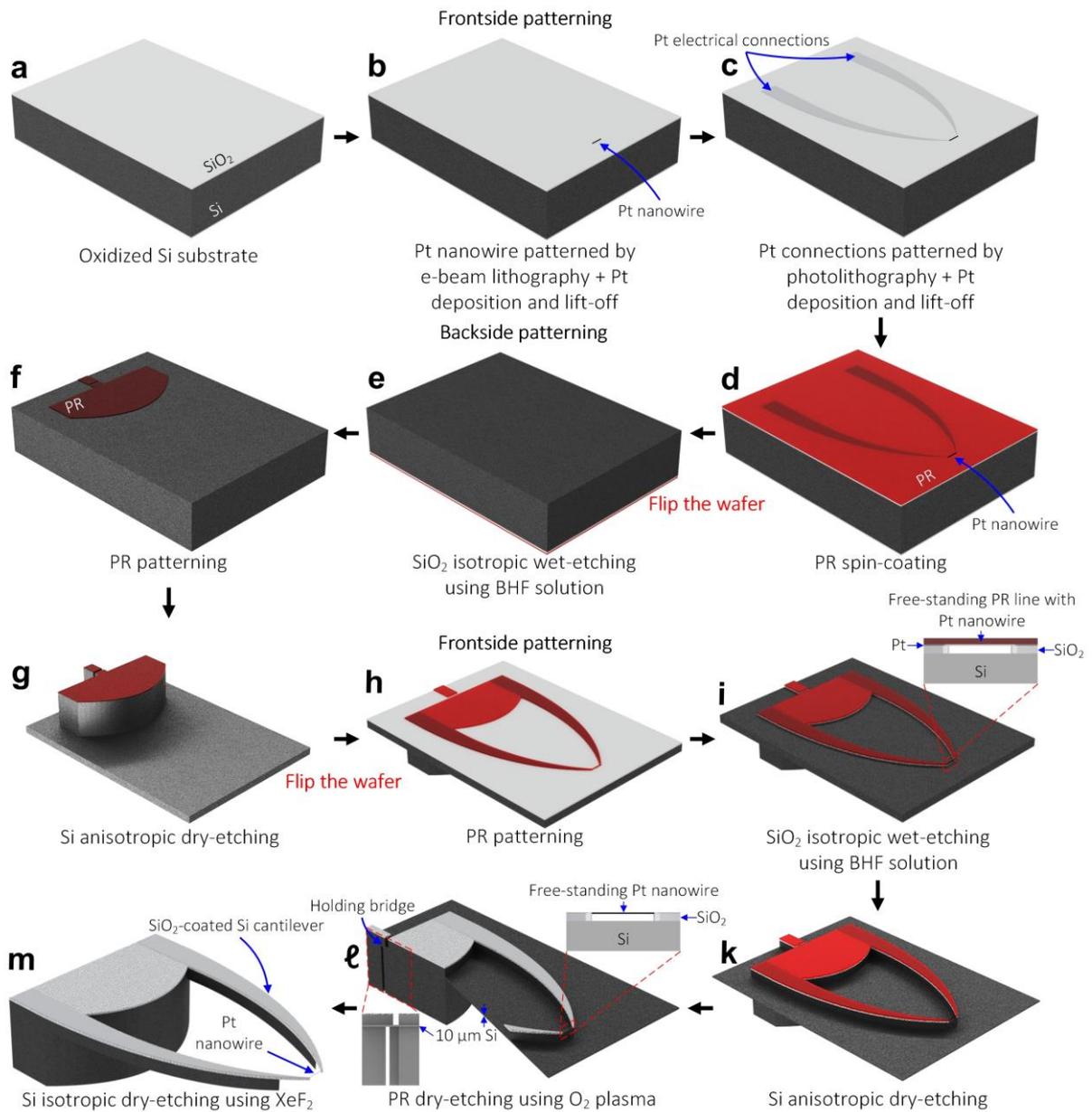

**Figure 1** Fabrication process for patterning free-standing Pt nanowires. (**a**) Wet thermal oxidation of a Si wafer. (**b**) Patterning a Pt nanowire using e-beam lithography. (**c**) Frontside patterning Pt connections to the Pt nanowire using standard photolithography. (**d**) Spin-coating of photoresist (PR) on the frontside of the patterned wafer, and (**e**) wet etching of $SiO_2$ layer on its backside using a BHF solution. (**f**) Backside patterning a PR structure of the device base, followed by (**g**) deep dry etching of Si. (**h**) Frontside patterning a PR structure of the support cantilevers. (**i**) Wet etching of $SiO_2$ using a BHF solution, resulting in a free-standing PR line

Free-standing Pt nanowires

with the Pt nanowire. (**k**) Dry etching of Si, followed by (**ℓ**) dry etching of PR using $O_2$ plasma at low power. (**m**) Isotropic dry etching of Si using $XeF_2$ for self-releasing of the device.

**Patterning Pt nanowires using electron beam lithography**

An e-beam lithography (EBL) system operating at 100 kV (Raith EBPG 5150, Raith GmbH, Germany) was used to pattern Pt nanowires on the surface of oxidized Si wafers (Figure 1b). These wafers were prepared by wet thermal oxidation of conventional (100) 4-inch silicon (Si) wafers (385 µm thick, Okmetic, Finland) (Figure 1a). Prior to the sputtering of Pt, a thin titanium (Ti) layer of ~13 nm thickness was sputtered in order to improve the adhesion of these patterned Pt nanowires. The choice of Ti for the adhesion layer is beneficial here because it can be easily removed together with $SiO_2$ layer in a buffered hydrofluoric acid (BHF) solution, thus leaving free-standing pure Pt nanowires. Figure 2 shows the high-resolution scanning electron microscopy (HR-SEM) images of a Pt nanowire fabricated on the surface of an oxidized Si wafer. A well-defined Pt nanowire was obtained with dimensions matching the specifications (~300 nm width, ~70 µm length, ~100 nm thickness). The pattern was expanded slightly at the wire tips to facilitate the electrical connection

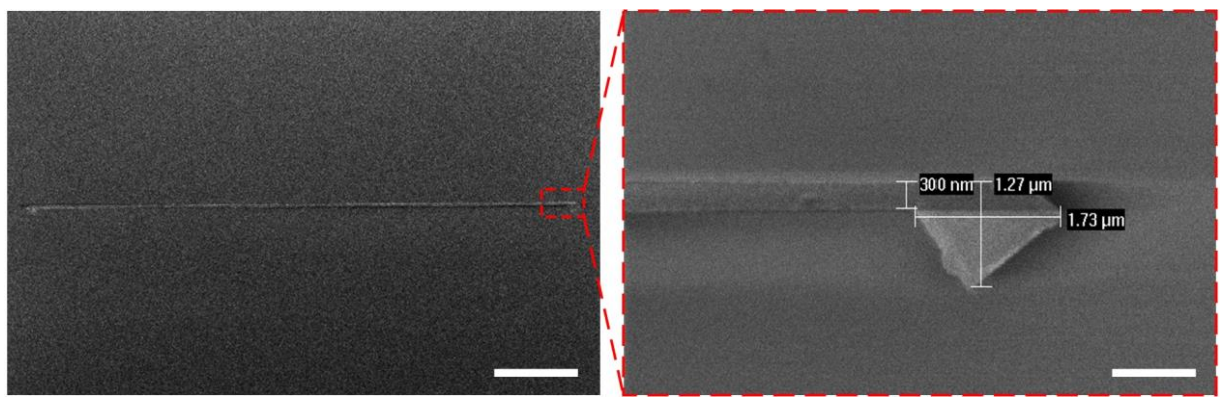

**Figure 2** Top-view HR-SEM image (scale bar: 10 µm) of a Pt nanowire patterned on the surface of an oxidized Si wafer, with a close-up image of the tip of the wire which is expanded slightly to facilitate the connection with the Pt micropattern (scale bar: 1 µm).

Free-standing Pt nanowires

**Patterning Pt connections to the Pt nanowires**

For the electrical connection to the Pt nanowire, Pt micropatterns (termed Pt connections) were fabricated by combining standard photolithography with a lift-off process (Figure 1c). Figure 3 shows the optical microscopy images of Pt connections patterned on the surface of an oxidized Si wafer. It should be noted that the precision of the overlay of the Pt connections with the Pt nanowire is crucial in this step as any misalignment between these structures can disrupt the electrical connection with the Pt nanowire.

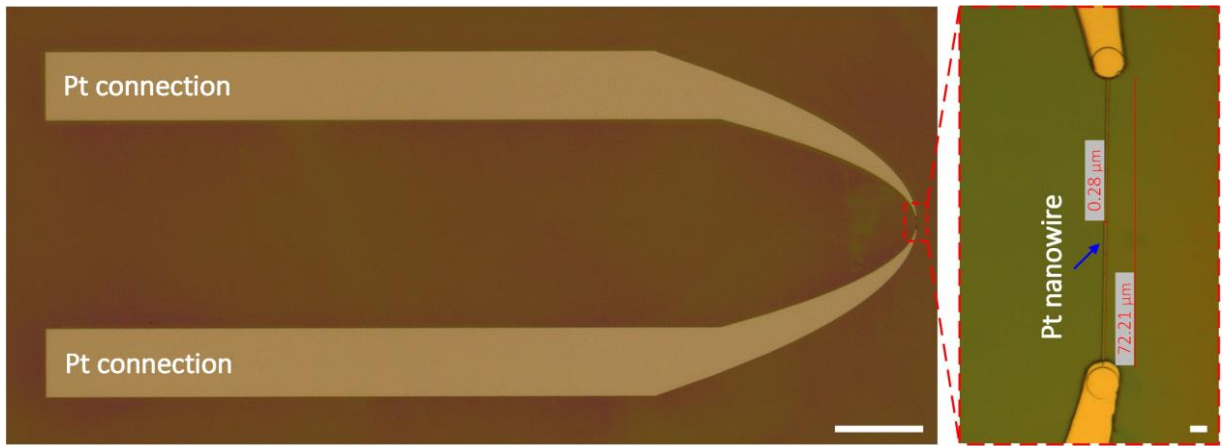

**Figure 3** Optical microscopy image (scale bar: 500 µm) of Pt connections, with a close-up image at the Pt nanowire location (scale bar: 5 µm).

**Backside patterning of the device base using dry etching of Si**

Prior to the backside patterning of the wafer, its frontside was covered with a PR layer (Figure 1d). The wafer was then immersed in a BHF solution in order to completely remove the $SiO_2$ layer on the backside (Figure 1e), while the $SiO_2$ layer on the frontside containing the patterned Pt structures remained protected by the PR coating.

Subsequently, a PR structure of the device base was patterned on the backside of the wafer using a standard photolithography process (Figure 1f). The patterned PR structure was hard-baked at 120°C for 10 min to harden the PR areas before conducting the etching of Si in an inductively coupled plasma (ICP) deep reactive ion etching (DRIE) instrument (SPTS Pegasus system, UK), using the standard Bosch process (Figure 1g). Figure 4 shows the HR-SEM

Free-standing Pt nanowires

images of the device base after the dry etching process. It is worth mentioning that a negatively tapered profile was obtained after the deep Si etching. This needs to be taken into account when designing the holding bridge for the self-releasing of the device (Figure 1ℓ).

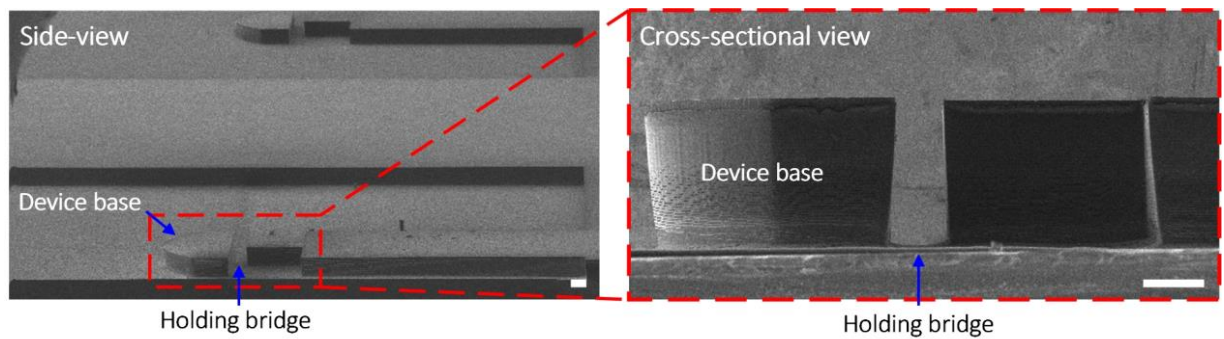

**Figure 4** Side-view and cross-sectional HR-SEM images (scale bar: 200 µm) of backside patterning of the device base using dry etching of Si.

**Frontside patterning of the device**

Figure 5 shows the optical microscopy images of a PR structure patterned on top of the Pt structure. Also in this case, the alignment of the patterned PR structure with the Pt structure needs to be precise so that the PR structure completely covers the Pt structure, especially at the Pt nanowire location where it is covered by a PR line, as shown in the close-up image (Figure 5). This ensures that the Pt structure is not damaged during the subsequent patterning of the cantilevers by wet etching and dry etching processes (Figures 1i and 1k).

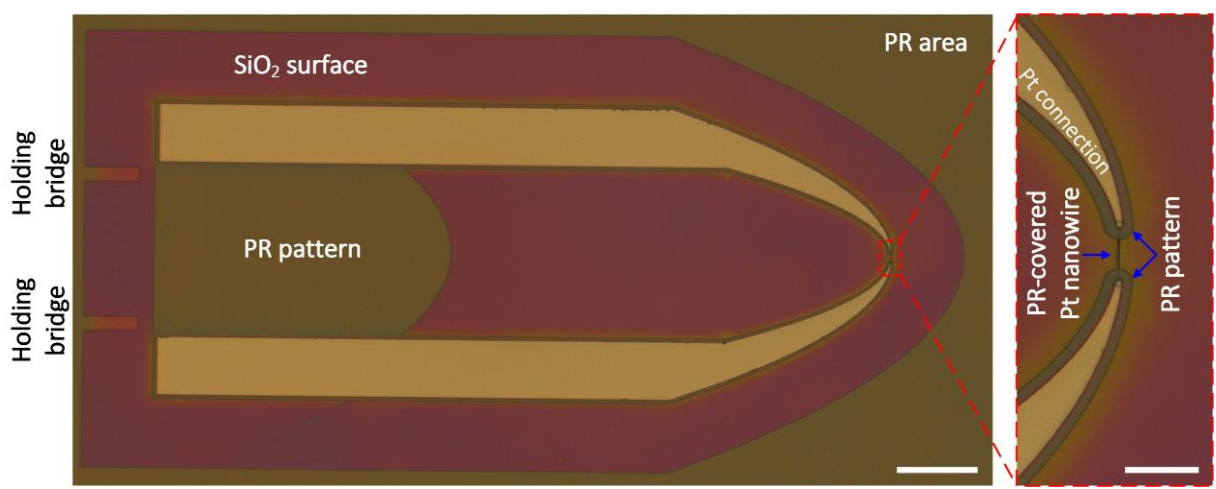

Free-standing Pt nanowires

**Figure 5** Optical microscopy image (scale bar: 500 µm) of a PR structure patterned on top of the Pt structure, with a close-up image at the Pt nanowire location (scale bar: 100 µm).

For releasing the PR line, the patterned wafers were immersed in a BHF solution for 30 min. As a result, the $SiO_2$ under the PR line was etched, thus leaving the free-standing PR line with the Pt nanowire stuck to it (Figures 1i and 6b). Since both PR and Si are hydrophobic any liquid trapped between the PR line and the Si surface was removed quickly and easily when spin-drying the wafers. Importantly, this resulted in no damage to the free-standing PR line supporting the Pt nanowire.

Figure 6a shows the side-view HR-SEM image of the support cantilevers after dry etching of Si from the frontside of the wafer. This etching process needs to be stopped when the thickness of the remaining Si membrane (Figure 1k) is down to approximately 10 µm. Etching through the Si-layer can lead to a leak of cooling gas from the backside, thus terminating the etching process. Crucially, further etching without cooling can result in burning of the Pt line, and hence a breaking of the Pt nanowire.

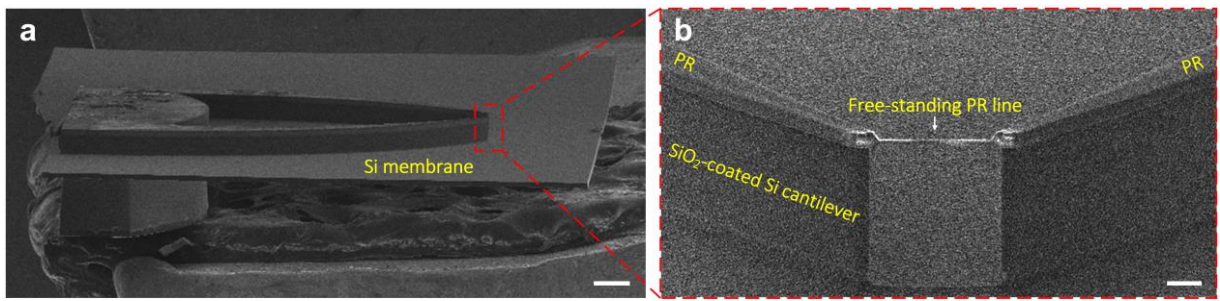

**Figure 6** (**a**) Side-view HR-SEM image (scale bar: 500 µm) support cantilevers after dry etching of Si from the frontside of the wafer. Note that the slight damage visible on the top surface of the device base was caused by handling during the SEM inspection. (**b**) Cross-sectional HR-SEM image (scale bar: 20 µm) of a free-standing PR line with the Pt nanowire.

To remove the PR covering the Pt nanowire, reactive $O_2$ plasma etching was used (Figure 1ℓ). This needs to be done gently at low power to avoid burning the PR line, and thereby breaking the Pt nanowire. The removal of PR was conducted before releasing the device because it turned out that the PR line became brittle after the dry etching process (Figure 1k). This resulted in frequent damage of the PR line during the releasing which then also affected the Pt nanowire.

**Isotropic dry etching of Si using XeF$_2$**

Figure 7 shows the HR-SEM images of a fabricated device consisting of a Pt nanowire free-standing between two SiO$_2$ beams supported on Si cantilevers (Figure 1m). After isotropic dry etching of Si in XeF$_2$, the remaining Si membrane was completely etched, forming two free-standing SiO$_2$-coated Si cantilevers (Figure 7b). It should be noted that the Si underneath the Pt nanowire as well as the Si at the tip of two cantilevers were also etched, thus resulting in the Pt nanowire free-standing on SiO$_2$ beams (Figure 7c).

Thanks to special design of the device holding bridge (Figure 1ℓ), the final etching step also served to self-release the device from the wafer. The holding bridge also has a remaining Si layer of ~10 μm and is thus etched away in XeF$_2$. This self-releasing procedure has proven necessary and important since it appeared that breaking the device led to frequent failures of the Pt nanowire (presumably due to the vibrations of the cantilevers). As confirmed in the close-up images, the resulting free-standing Pt nanowire has a width of ~300 nm and a length of ~70 μm.

Free-standing Pt nanowires

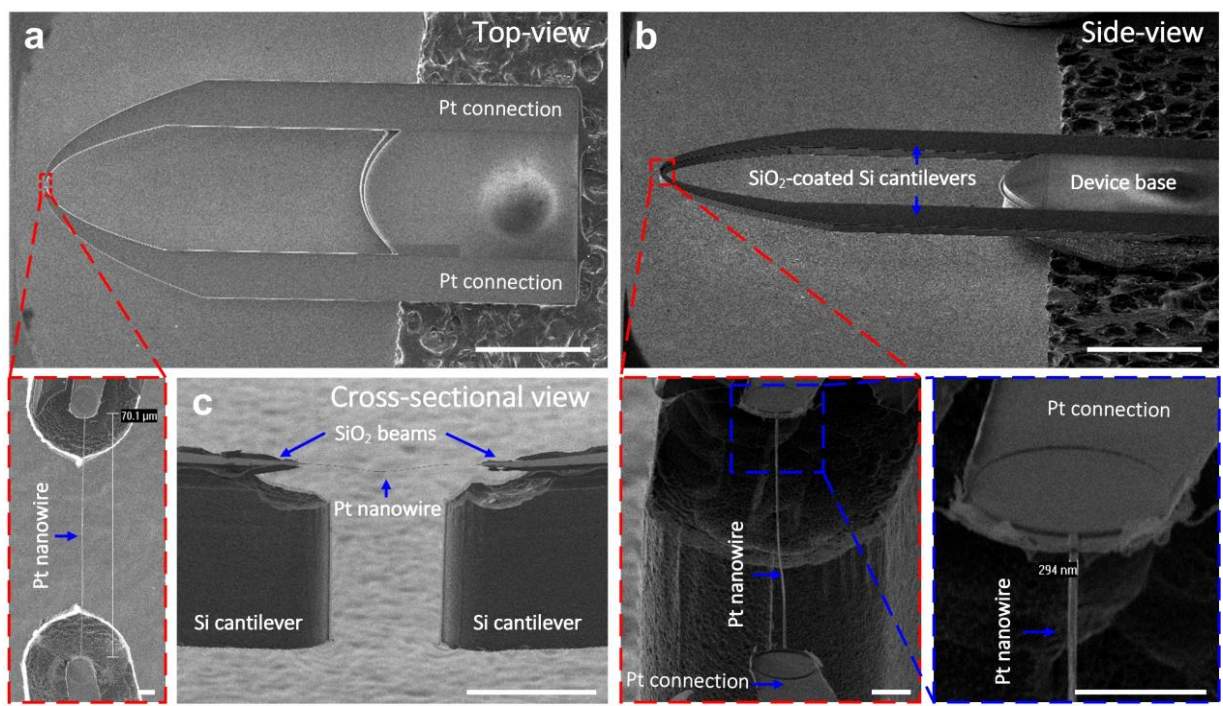

**Figure 7** (**a**) Top-view and (**b**) side-view HR-SEM images (scale bar: 1 mm) with close-up images (scale bar: 5 µm) of a fabricated device consisting of a free-standing Pt nanowire (~300 nm width, ~70 µm length, ~100 nm thickness). (**c**) Cross-sectional HR-SEM image (scale bar: 50 µm) of a fabricated Pt nanowire hanging between two $SiO_2$ beams supported on Si cantilevers.

**Batch size and fabrication yield**

With our mask design, each 4-inch wafer contains 150 devices. Typical yields in the trial fabrication processes performed so far were about 50% - 70% (~70 – 100 functional devices per wafer). A limiting factor for the fabrication yield was the manual handling of the self-released devices by tweezers after dry-etching. We believe that the fabrication yield can be increased further by improving the device handling and by further optimizing the fabrication process steps, especially regarding the uniformity of the dry-etching steps over the entire wafer. Additionally, it should be mentioned that also even thinner wires with widths of 200 nm and 100 nm were successfully produced with this process. However, in these cases the fabrication

Free-standing Pt nanowires

yield was impractically low and the wires turned out to be not robust enough. We therefore did not pursue the production of wires thinner than 300 nm further.

**Performance of the fabricated devices used as thermal anemometer probes**

A typical initial cold resistance for the nanowire is 820Ω, but this value was seen to drop significantly when the wire was first heated up. Annealing the nanowire with incrementally increasing currents up to ~1 mA reduced the resistance to $R_w \approx 740\Omega$, and this value was found to be stable over repeated heating cycles with comparable currents. The annealing was performed in the actual experiment with a weak flow of either air or $SF_6$ gas. While the cold resistance appeared stable after shorter times already, we typically annealed over several hours in order to avoid any spurious drift in the subsequent tests. By measuring the wire resistance in a temperature controlled environment we determined the temperature coefficient of resistivity to be $\alpha_{20°C} = 0.0021$ K$^{-1}$.

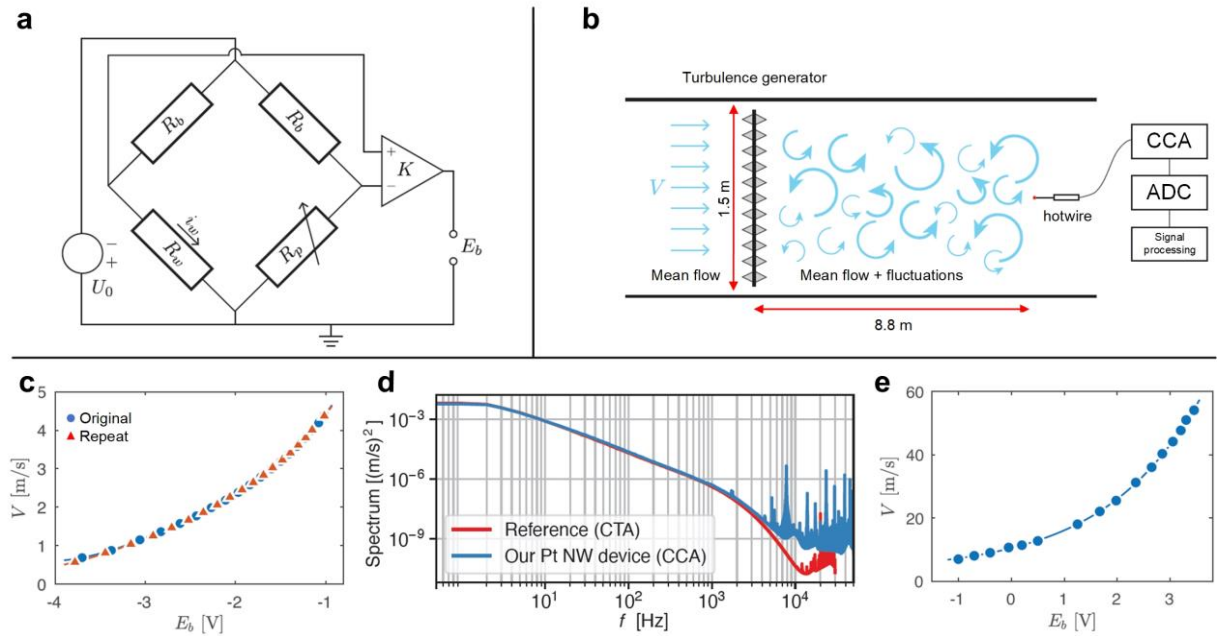

**Figure 8** (**a**) Sketch of the CCA circuit. (**b**) Measurement setup in the VDTT. (**c**) Calibration results in the VDTT along with a repeat taken several hours later. (**d**) Velocity power spectrum as a function of frequency $f$ recorded with our device (in constant current mode) compared to a

Free-standing Pt nanowires

commercial reference probe run by a constant temperature anemometer (CTA). (**e**) Calibration results in air at room temperature.

The nanowires were operated in a bridge circuit (see figure 8a) and tested in the Variable Density Turbulence Tunnel (VDTT) in Göttingen[24] in a gaseous sulfur hexafluoride ($SF_6$) environment up to extremely high Reynolds numbers (see schematic in figure 8b). Note that the purpose of using $SF_6$ here is to reduce the kinematic viscosity compared to e.g. air which allows to reach high *Re* while keeping flow velocities moderate. This effect can be enhanced by pressurizing the tunnel up to 15 bar. Further details of the setup and operating conditions are given in the Materials and Methods section. In order to calibrate the sensor output voltage $E_b$ against the fluid velocity, a time average of $E_b$ was recorded for several settings of the tunnel velocity *V* in non-turbulent conditions. In order to gauge a potential drift of the bridge voltage, the calibration was performed both before and after a measurement series that spanned several hours. The calibration results are presented in Figure 8(c). There is a clear and monotonic trend between $E_b$ and *V* that can be captured very accurately over the full velocity range by fitting a fourth-order polynomial (indicated by the lines), which is a standard procedure for hotwire measurements[25]. Importantly, calibration results before and after the measurement series are almost indistinguishable, indicating that the drift of the sensor is negligible over an operation period of several hours. As an additional validation, we compare the energy spectra of the fluctuating velocity *v(t)* measured by our probe to results obtained using a standard probe (length 450 µm, diameter 2.5 µm, Dantec Dynamics custom design) as a reference in Figure 8(d). These measurements were taken in 2 bar $SF_6$ with a mean flow velocity $V = 3.75$ m s$^{-1}$. The Taylor Reynolds number was $Re_\lambda = 990$ and the viscous length scale $\eta \sim 63$ µm. Generally the spectra agree very closely between the two sensors which also manifests in the fact that the velocity variances (i.e. the integral of the spectra) differ only by about 1%, which is on the order of the discrepancy expected as the probes are not located in the exact same location. The collapse of the spectra up to a frequency $f \sim 1$ kHz is in particular remarkable since the reference was operated in a constant temperature mode, which offers superior temporal resolution characteristics to the constant current mode employed to operate our wire here. The slightly elevated noise level at very high frequencies on the order of 10 kHz for our nanowire is also a result of the rather basic circuitry and components employed for these first tests. Further, there were no issues operating the wires at pressures up to 15 bar in $SF_6$, at which the gas density is more than $1/10^{th}$ that of water at room temperature. To test whether the wires also perform well in other fluids and at larger flow speeds, we additionally operated the wire in air at room temperature. In this case the flow was generated by pressurized air exiting a nozzle. Also here, the data can be very well represented by a monotonically increasing fourth-order polynomial across the full range of 5 m s$^{-1}$ $\leq V \leq$ 55 m s$^{-1}$. The wire was able to withstand the dynamic pressure at the highest velocities without any problems (Figure 8e).

**CONCLUSION**

In summary, we report a robust fabrication method, combining e-beam lithography with wet etching and dry etching processes, for patterning free-standing Pt nanowires used as thermal anemometer probes for turbulence measurements. With precise control of the dry etching processes, Pt nanowires (~300 nm width, ~100 nm thickness) with a length of 70 µm have been successfully released free-standing between two $SiO_2$ beams that are supported on Si cantilevers. A critical aspect is the design of the holding bridge, which ensures a safe and gentle release of the device without damaging the wires. Further, limiting the use of e-beam lithography to the patterning of the Pt nanowires renders the process cost and time efficient. These benefits far outweigh the additional complication arising from the resulting need to align e-beam and optical lithography patterns with high accuracy. The operational tests have confirmed that the wires are suitable for turbulence measurements in different working media and at a high dynamic pressures.

Free-standing Pt nanowires

Further characterizations and developments regarding the circuitry, in particular the implementation of a CTA capable of handling the relatively high wire resistances, are necessary to exploit the full potential. However, it is already clear that the nanowire design presented here holds a lot of promise regarding several aspects: (i) A more slender wire allows to use shorter wire lengths without being compromised by end-conduction effects; (ii) Smaller sensing elements are expected to improve the frequency response of the anemometer even if the wire is operated in a constant temperature mode[26]; (iii) Due to its very small thermal inertia, the wires can yield sufficient frequency resolution for many flow cases even when operated in constant-current mode, as our preliminary results here prove. This eliminates the need for a feedback loop, thereby simplifying the circuitry significantly; (iv) The quasi circular shape of the sensing element is expected to avoid an unwanted pitch sensitivity of the sensor. We aim to explore and quantify these benefits in the future in an effort to push the limits for highly resolved high *Re* turbulence measurements.

**MATERIALS AND METHODS**

**Wet thermal oxidation of Si wafers**

Conventional (100) 4-inch Si wafers (385 µm thick, Okmetic, Finland) with a thick thermal oxide layer of approximately 2 µm were prepared by wet thermal oxidation (Figure 1a). Prior to the wet thermal oxidation process, all the Si wafers were cleaned to prevent cross-contamination[27]. Subsequently, the Si wafers were loaded into a high temperature tube furnace (Model 287, TEMPRESS), using a quartz carrier to implement the wet oxidation at 1150°C for 12 h. During the oxidation process, the flow rate of a mixture of water vapor and nitrogen gas was fixed at 2 ℓ min$^{-1}$. The ramping and cooling rates were set at 10°C min$^{-1}$ and 7°C min$^{-1}$, respectively.

**Patterning Pt nanowires using electron beam lithography**

Free-standing Pt nanowires

Prior to the e-beam writing, positive resist (NANO™ 950PMMA Series Resists in Chlorobenzene, MicroChem, US) was spin-coated over the surface of the oxidized Si wafers at 2500 rpm for 45 s, followed by baking at 165°C for 2 min. Subsequently, an e-beam lithography (EBL) system with a 100-kV (Raith EBPG 5150, Raith GmbH, Germany) was used to write the nanowire pattern into the resist layer. The written wafers were then developed in a developer solution (MIBK-IPA mixture) in 90 s, followed by rinsing with deionized (DI) water using a quick dump rinser and spin-drying with nitrogen ($N_2$).

A titanium (Ti) layer of ~13 nm and a platinum (Pt) layer of ~100 nm were sputtered over the patterned wafers using an ion-beam sputtering system (home-built T'COathy system, MESA+, NanoLab)[28]. The sputtering processes were performed at 200 W and at a pressure of $6.6 \times 10^{-3}$ mbar which was adjusted using an argon (Ar) flow. Subsequently, the wafers were immersed in acetone with sonication to perform the lift-off process. After rinsing the wafers with DI water and spin-drying with $N_2$, the fabrication of Pt nanowires patterned on the surface of the oxidized Si wafers was finished (Figure 1b).

**Patterning Pt connections to the Pt nanowires**

A positive photoresist (PR) layer (OiR 907-17i, Fujifilm, Japan) was spin-coated over the wafer surface at 4000 rpm for 45 s, followed by baking at 95°C for 1 min. A photo-mask made of quartz containing inverted chromium (Cr) patterns connected to the patterned Pt nanowires was fabricated in-house by using a mask-making system (DWL 2000 Laser Lithography System, Heidelberg Instruments, Germany). The exposure process was performed by using a mask alignment system (EVG620, EV Group, Austria) for 5 s at an intensity of 12 mW cm$^{-2}$ in hard contact mode. Thereafter, the wafers were post-baked at 120°C for 1 min, followed by developing in a OPD4246 solution for 1 min, rinsing with DI water, and drying with $N_2$. A Ti layer of ~6 nm and a Pt layer of ~100 nm were sputtered over the patterned wafers using the T'COathy system. The lift-off process was conducted in acetone with sonication, followed by

Free-standing Pt nanowires

rinsing the wafers with DI water. After spin-drying with $N_2$, the fabrication of Pt connections to the Pt nanowires was completed (Figure 1c).

**Backside etching of the thermal oxide layer**

The patterned surface of the oxidized Si wafers was covered with a PR layer (OiR 908-35, Fujifilm, Japan) by spin-coating at 2000 rpm for 45 s, followed by baking at 95°C for 3 min (Figure 1d). The wafers were then immersed in a buffered hydrofluoric acid (BHF) solution for 30 min to remove the $SiO_2$ layer completely (etch rate of ~68 nm min$^{-1}$) on their backside (Figure 1c).

**Backside patterning of the device base using dry etching of Si**

After removing the PR layer in acetone, cleaning with DI water and drying with $N_2$ gas, the backside of the wafers was spin-coated with a PR layer (OiR 908-35, Fujifilm, Japan) at 2000 rpm for 45 s, followed by baking at 95°C for 3 min. A photo-mask containing a Cr pattern of the device base was used for the exposure process, which was performed by using the mask alignment EVG620 system for 15 s at an intensity of 12 mW cm$^{-2}$ in hard contact mode. The alignment with the frontside Pt structures was performed using the bottom alignment with cross-hair mode. Thereafter, the wafers were post-baked at 120°C for 1 min, followed by developing in the OPD4246 solution for 3 min, rinsing with DI water, and drying with $N_2$. Subsequently, the wafers were baked at 120°C for 10 min to harden the remaining PR areas for further backside etching of the Si (Figure 1f).

The etching of Si was conducted in an inductively coupled plasma (ICP) deep reactive ion etching (DRIE) instrument (SPTS Pegasus system, UK), using the standard Bosch process with 105 cycles (0.6 s deposition of $C_4F_8$, 1.75 s etching of Si by $SF_6$) (Figure 1g). After the deep Si etching, the wafers were immersed in a 99% nitric acid ($HNO_3$) solution for 30 min to completely remove the PR layer and any other residues.

**Frontside patterning of the device**

Subsequently, the wafers were flipped and their frontside was spin-coated with a positive PR layer (OiR 907-17i, Fujifilm, Japan) at 4000 rpm for 45 s, followed by baking at 95°C for 1 min. A photo-mask containing Cr pattern of support cantilevers was used for the exposure process was performed by using the mask alignment EVG620 system for 5 s at an intensity of 12 mW cm$^{-2}$ in hard contact mode. The wafers were then post-baked at 120°C for 1 min, followed by developing in the OPD4246 solution for 1 min, rinsing with DI water, and drying with $N_2$. Subsequently, the wafers were baked at 120°C for 10 min to harden the PR layer (Figure 1h).

*Releasing of the PR line with the Pt nanowire*

The patterned wafers were then immersed in the BHF solution for 30 min to remove the unprotected $SiO_2$ layer completely. Since the PR line covering the Pt nanowire at the tip of the cantilevers has a small width of approximately 3 µm, etching in the BHF solution for 30 min resulted in a complete removal of $SiO_2$ under the PR line and Ti under the Pt nanowire. As a result, the PR line with the Pt nanowire stuck to it got released in this step (Figure 1i).

*Patterning support cantilevers using dry etching of Si*

The wafers were then etched in the SPTS Pegasus system using the fine-etching process with 90 cycles (Figure 1k) until the remaining Si layer reached a thickness of approximately 10 µm.

*Etching of the PR line using $O_2$ plasma*

To remove the PR covered the Pt nanowire, oxygen ($O_2$) plasma etching was performed in a parallel plate reactive ion etching system (home-built TEtske system, MESA+, NanoLab) at

Free-standing Pt nanowires

wafer-level, 10 mTorr, and 25 W for 20 min. A low power etching was used in order to not break the Pt nanowire during the etching of PR (Figure 1ℓ).

**Isotropic etching of Si using XeF$_2$**

For the final patterning of the cantilevers and for releasing the devices, the wafers were put in a gas phase Xactix XeF$_2$ E1 system (etching time per cycle: 30 s, temperate: 35°C, pressure: 3000 mTorr), so that the Si was isotropically etched by xenon difluoride (XeF$_2$, etching rate of ~1 μm) (Figure 1m). This resulted in an etching through of the remaining Si layer, thus forming two free-standing SiO$_2$-coated Si cantilevers. The Si underneath the Pt nanowire and the Si at the tip of two cantilevers were also etched, thus resulting in the Pt nanowire free-standing on SiO$_2$ beams. The device was also self-released after this etching step thanks to the special design of the device holding bridge.

**Electrical connection to the device using a silver conductive glue**

For the electrical connection, the fabricated device was mounted on the prongs of a commercial probe holder (Dantec Dynamics A/S, Denmark) using a silver conductive glue (Figure S2). In order to cure the glue, the device mounted probe was baked in an oven at 120°C for at least 15 min.

**Testing the fabricated devices used as thermal anemometer probes**

To operate the nanowire, we used a constant current anemometer (CCA) circuit as sketched in Figure 8a. Here, the device was placed in a bridge that features large ballast resistances $R_b$ = 12 kΩ at the top of both arms. Since $R_b$ ▯ $R_w$, this ensures that the wire current $i_w$ remains essentially constant, even as $R_w$ changes slightly. With the nanowire exposed to the flow, we adjust the bridge voltage $U_0$ until the desired overheat ratio $a = R_w = R_{w;20°C}$ is reached with typical values of $a = 1.2 - 1.4$ corresponding to wire overheat temperatures of 100°C – 200°C.

Free-standing Pt nanowires

The resistance $R_p$ is chosen such that the bridge is balanced at working conditions. The bridge voltage is then proportional to small differences in $R_w$ that come about as the time varying cooling by the flow changes the wire temperature slightly. Amplified by a factor $K = 100$ via an instrumentation amplifier, the bridge voltage $E_b$ is recorded as the output parameter of the CCA using an analogue-digital converter (ADC). A calibration and additional signal processing (*e.g.* filtering) as required finally yield the desired measurement of the fluctuating fluid velocity. The nanowire was tested in the VDTT in Göttingen described elsewhere[24]. The device was placed in the freestream behind an active turbulence generating grid as sketched in Figure 8b. The grid triggers turbulent motion in the fluid such that the fluid velocity *v(t)* at the hotwire location fluctuates in time around its mean *V*. For the present set of measurements, the VDTT was operated at a pressure $p = 2$ bar with $SF_6$ at a temperature of 21°C as the working medium. The overheat ratio was set to $a = 1.24$ and the wire current was $i_w = 0.622$ mA.

## ACKNOWLEDGEMENTS

This work is supported by the Max Planck-University of Twente Center for Complex Fluid Dynamics and by the Netherlands Organisation for Scientific Research (NWO) Gravitation programme funded by the Ministry of Education, Culture and Science of the government of the Netherlands.

## COMPETING INTERESTS

The authors declare no conflict of interest

## SUPPLEMENTARY INFORMATION

Supplementary information accompanies the manuscript on the Microsystems & Nanoengineering website: http://www.nature.com/micronano.